\documentstyle[12pt,epsf]{article}
\def\reference{\parskip 0pt\par\noindent\hangindent 0.5 truecm}

\begin{document}
%
% Title
% Capitalise the title normally - do not use ALL CAPS.
%
\title{The Warm Ionized Medium in Spiral Galaxies: A View from Above}
%

% Authors
% Here comes the author(s) of the paper. Please add the appropriate author
% names for your paper and indicate within the $^...$ the number(s)
% which corresponds to the institute(s) of each author. In this example
% the second author has two institutional affiliations.
% Add or remove authors as required, maintaining the \and syntax between
% each author, but no \and after the last author.
% **** IMPORTANT: Leave the closing curly bracket line as is. ******

\author{Ren\'e A.M. Walterbos $^{1}$ 
} % IMPORTANT: leave this curly bracket as the first character of this line.

% Date - leave this blank.
\date{}
\maketitle

% Institutions
% Here fill in your institute name(s) and address(es)
% The number in $^...$ indicates the author number.  For example
{\center
$^1$ NMSU, Astronomy Department, MSC 4500, Box 30001, Las Cruces, NM 88003, USA\\rwalterb@nmsu.edu\\[3mm]
}

% Abstract
% Simply place your abstract between the \begin{abstract} and
% \end{abstract} commands.
%
\begin{abstract}
% Place the abstract here.
  The Warm Ionized Medium (WIM), also refered to as Diffuse Ionized
  Gas, contains most of the mass of interstellar medium in ionized
  form, contributing as much as 30\% of the total atomic gas mass in
  the solar neighborhood. The advent of CCDs has enabled unprecedented
  study of this medium in external galaxies, probing a variety of
  environments.  In particular, we can derive the morphology of the
  WIM, its distribution across disks, and the correlation with other
  Population I material.  Spectroscopy of the WIM makes it possible to
  test various ionization models. I will review here our current
  understanding of the properties of the WIM in spiral galaxies. A
  perhaps unexpected result is that the H$\alpha$ emission from the
  WIM contributes about 40\% of the total observed H$\alpha$
  luminosity from spirals. This places severe constraints on possible
  sources of ionization, since only photo ionization by OB stars meets
  this requirement. Spectroscopic measurements of forbidden line
  strengths appear in reasonable agreement with photo ionization
  models. It is not yet clear if the Lyman continuum photons that
  ionize the WIM are mostly from OB stars located inside traditional
  HII regions, or from field OB stars.
\end{abstract}

{\bf Keywords:}
% Place keywords here.  PASA uses the standard list of subject 
% headings adopted by The Astrophysical Journal and available from URL:
%   http://www.noao.edu/apj/keywords96.html
Galaxies: Local Group, spiral, ISM: HII regions, bubbles, Ultraviolet: stars
% A formatting command to add space between the author list and the body
% of the paper when printed. This spacing may be changed as desired.
\bigskip

%
% Body of paper
%

\section{Introduction}

% Place contents of first section here.

Warm Ionized Medium (WIM), or Diffuse Ionized Gas, is the dominant
component of the ionized Interstellar Medium (ISM) in disk galaxies.
While the H$\alpha$ emission from this component is 10 to 1000 times
fainter than for traditional HII regions, and the gas has low density
($\rm n_e \sim 0.2$ cm$^{-3}$), its large volume filling factor and
spatial extent imply that the mass of the WIM easily surpasses that
contained in traditional HII regions or in the hot gas in the ISM.
Understanding the heating and ionization mechanism for the WIM is a
major challenge to models of the ISM.  In external galaxies we can
determine the overall distribution and morphology of the WIM across
galactic disks, its correlation with other ISM phases, and the
variation in its properties with Hubble type and star formation rate.
In addition, we can test ionization models for the WIM through
spectroscopy, and through determining the relation between WIM and
ionizing stars. In this paper we will review results for galaxies that
are not edge-on; see Rand (this volume) for results on edge-on
systems.

\section{The Warm Ionized Medium in Disk Galaxies}

% Place contents of next section here.
Emission line imaging with CCDs on even modest size telescopes is an
excellent method for studying the WIM in galaxies, provided care is
taken in flat fielding and subtraction of continuum light. Imaging
with Fabry-Perot systems is another fruitful observational approach,
as discussed by Bland-Hawthorne (this volume).  Most imaging studies
have focused on H$\alpha$, sometimes including [NII](6548+6583\AA),
and on [SII](6716+6731\AA) emission lines (e.g. Walterbos \& Braun
1992, 1994, Hoopes et al. 1996, Ferguson et al. 1996a,b). The
H$\alpha$ intensity one observes is directly proportional to the
Emission Measure (EM), the integral of the electron density {\it
  squared} along the line of sight. For ionized gas at 10,000K, $\rm
1\ Rayleigh = 5.6 \times 10^{-18}\ erg\ cm^{-2}\ s^{-1}\ arcsec^{-2} =
2.8\ pc\ cm^{-6}.$ The WIM is too diffuse to obtain density
information from the ratios of forbidden lines such as the [SII]
doublet.  Thus we have no direct probe of the electron density, hence
column density of ionized gas, in observations of external galaxies.
For the Galactic WIM, this information does exist through observations
of pulsar dispersion measurements (see e.g.  Kulkarni \& Heiles 1988
for a review).

\begin{figure}
\centering \leavevmode
\epsfxsize=1.0\textwidth
\epsfbox{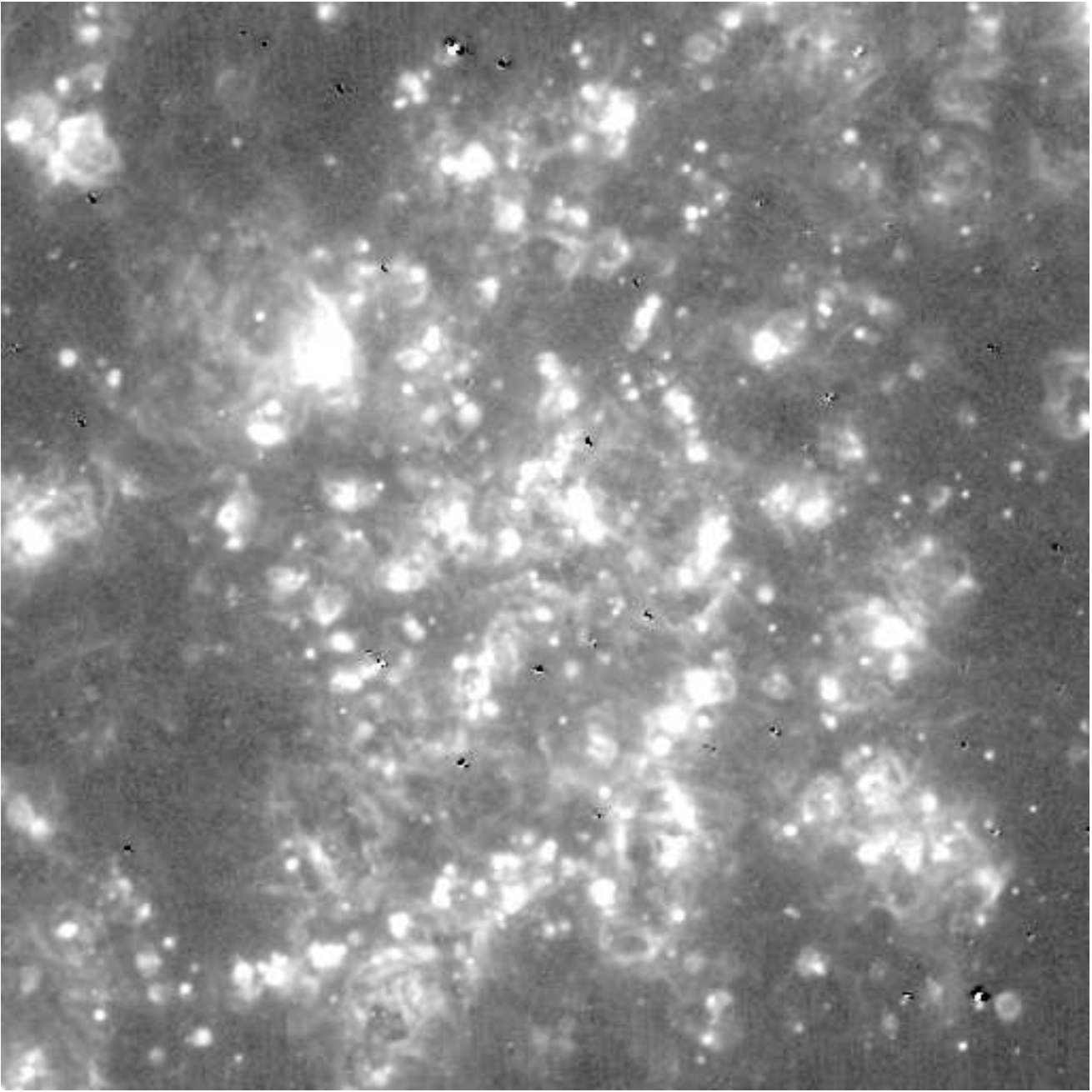}
\caption{A 5-hour exposure in H$\alpha + $[NII] of the nearby spiral 
  M33 obtained with the Burrell Schmidt telescope at Kitt Peak.  The
  image shows the central 4.2 by 4.2 kpc$^2$. The brightest HII region
  in M33, NGC 604 is located just left and above the middle. The grey
  scale saturates at an EM of 500 pc cm$^{-6}$. Continuum light has
  been subtracted. The WIM seems to cover almost the entire area of
  the disk not occupied by traditional bright HII regions (from
  Greenawalt 1997). }
\label{fig-1}
\end{figure}

We show an example of a deep H$\alpha$ image for the nearby spiral M33
in Figure~\ref{fig-1}. The WIM is brightest in regions with a high
surface density of bright HII regions, although sometimes WIM patches
or filaments are located far away from traditional HII regions (see
also Hunter et al. 1990, 1992). The WIM covers a large fraction of the
disk, here as much as 100\%. Its morphology is a combination of
diffuse emission and curved filaments, perhaps consistent with a
``dented sheet''.  Typical Emission Measures contributed by the WIM
reach up to as much as 50 pc cm$^{-6}$ in spiral arms, down to as low
as a few pc cm$^{-6}$ at the faintest levels so far detected. For
comparison, in the solar neighborhood the Reynolds layer has an
emission measure of about 5 pc cm$^{-6}$ perpendicular through the
disk.

H$\alpha$ imaging allows straightforward determination of a crucial
quantity: the {\it fractional} H$\alpha$ luminosity contributed by the
WIM in a galaxy.  A detailed method on how to separate the WIM
contribution from the total H$\alpha$ luminosity has been described by
Hoopes et al. (1996).  Results are now available for M31 (Walterbos \&
Braun 1994), NGC 253, NGC 300 (Hoopes et al.  1996), NGC 247, NGC 7793
(Ferguson et al.  1996a), NGC 55 (Hoopes et al. 1996, Ferguson et al.
1996b, M81, M51, and M33 (Greenawalt 1997). A surprising
result emerges: the contribution from DIG is 40 $\pm$ 10 \%,
irrespective of the star formation rate in all these galaxies.  Such a
result might be expected if the H$\alpha$ emission we observe from the
WIM were in fact scattered light from bright HII regions in galaxies.
However, the distinct spectral signature (see next section) and the
distinct morphology of the WIM that is discernible in nearby galaxies
(e.g. Walterbos \& Braun 1994) make this unlikely. If the extinction
is systematically less in the WIM compared to that in HII regions,
this number may be somewhat less.  This has only been addressed for
M31 (Walterbos \& Braun 1994, Greenawalt et al. 1997), where the
corrected WIM fraction probably remains at least 20 to 30\%.  The high
number for this fraction is relevant in that it forces us to accept
that OB stars have to be the dominant ionization mechanism. In
addition, the fact that this fraction is constant among galaxies
argues against a strong influence for an external ionizing source.
Instead, the constant fraction must be reflecting some fundamental
property of the ISM and the distribution of ionizing sources in
galaxies. Either the medium is similarly porous in galaxies with
widely different star formation rates per unit area, or the ratio of
{\it field} OB stars to total number of OB stars is similar in all
galaxies.

\section{Spectroscopy of the WIM and the Source of Ionization}

The ionization problem for the WIM has two aspects. First, the energy
requirements to keep the medium ionized are enormous, leading to OB
stars as the only viable candidates. But do photo ionization models
predict the correct spectrum for the WIM? Second, there is a transport
problem, in that the mean free path for Lyman continuum photons in
galactic disks is very small, typically less than a pc (this small
number implied the existence of ``Str\"omgren spheres'' in the ISM in
the first place). So how can the ISM be ionized at large distances
from OB stars? Do Lyman continuum photons leak from HII regions, or
are field OB stars responsible?

\begin{figure}
\centering \leavevmode
\epsfxsize=0.50\textwidth
\epsfbox{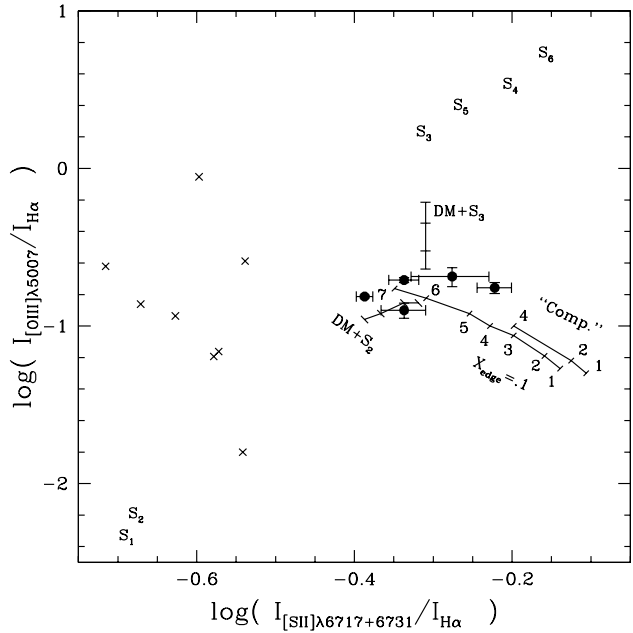}
 \caption{Line ratios for the WIM in M31  (filled dots) compared to 
   various ionization models. The crosses indicate bright HII regions
   in M31. The full drawn lines indicate loci of photo ionization
   models from Domg\"orgen \& Mathis (1994), for various diluted
   radiation fields. The ``S'' points refer to mixing layers models
   (Slavin et al. 1993) for various temperatures and mixing speeds.
   The photo ionization models give satisfactory agreement and only a
   modest contribution from mixing layers appears to be allowed (from
   Greenawalt et al. 1997).} 
\label{fig-2}
\end{figure}

A characteristic of the Galactic WIM is its high [SII] over H$\alpha$
emission line intensity ratio (e.g. Reynolds 1988). The WIM in other
galaxies shows the same behavior (e.g. Walterbos \& Braun 1992, 1994,
Rand et al. 1990, Ferguson et al. 1996b, Hoopes et al 1996).
Spectroscopic results by Greenawalt et al.  (1997; see
Figure~\ref{fig-2} here) for M31 and imaging results by Ferguson et
al. for NGC 55 (1996b) show [SII]/(H$\alpha +$[NII]) to increase with
decreasing H$\alpha$ intensity. In M31, [NII]/H$\alpha$ and
[OIII]/H$\beta$ do not
change. These results agree with photo ionization models of
Domg\"orgen \& Mathis (1994) who calculate various line ratios for
diffuse gas exposed to a strongly diluted radiation field.  Overall,
the WIM in M31 shows similar, but not identical, spectral
characteristics as the Reynolds layer. The differences (e.g. stronger
[OIII]/H$\beta$ in M31) indicate an overall less diluted radiation
field in M31 compared to the solar neighborhood, not surprising given
that for M31 the relatively bright WIM in the spiral arms was
observed.

Ferguson et al. (1996) imaged NGC 55 in [OII], detecting a smoothly
increasing [OII]/(H$\alpha$ +N[II]) ratio with decreasing H$\alpha$
intensity of the WIM. The WIM in M31 may show the same behavior
(Greenawalt et al. 1997). This trend appears not to be predicted
in the photo ionization models of Domg\"orgen \& Mathis (1994).
Greenawalt et al. (1997) also looked at the predicted line ratios for
mixing layer models (Slavin et al. 1993); it seems that a possible
contribution from mixing layers to line emission from the WIM is less
than 20\%.

In distinguishing which spectral type of OB stars are playing a role
in ionizing the WIM, knowledge of the ionization stage of Helium is
crucial, since only stars earlier than O8 can ionize
Helium in significant amounts. Early results for Galactic WIM at an
average EM of 30 pc cm$^{-6}$ (Reynolds \& Tufte 1995) indicated that
most of the Helium in the direction they studied had to be neutral.
This caused significant problems for photo ionization models, since
not enough ionizing radiation could be contributed by the
late-spectral type stars implied to be responsible for the ionization
of the WIM. More recently (see Reynolds, this volume), the He(5876\AA)
recombination line has been detected for Galactic WIM and in the halo
gas of NGC891 (Rand 1997), but it appears that He is not fully
ionized. Greenawalt et al. 1997) concluded that for relatively bright
WIM in M31 (at EM above 50 pc cm$^{-6}$), Helium appears to be fully
ionized. We could not derive information for WIM at lower intensity
levels. It is clear that further measurements of the He recombination
line are required in different environments.

\section{Can Field OB Stars Ionize the WIM?}

Given that it appears likely that the WIM is ionized by OB stars,
where do the Lyman continuum photons originate from: leakage from HII
regions, or field OB stars?  The first possibility agrees with several
observed characteristics of the WIM: the increase in forbidden line
strengths compared to the Balmer lines towards lower H$\alpha$
intensities, and the concentration of H$\alpha$ emission from the WIM
near HII regions.  Ferguson et al. (1996a) argue that leakage has to
occur because field OB stars may not be capable of contributing enough
ionizing photons. However, a more careful census of the field star
population is necessary.  An O star located in a low-density
environment will have a very large Str\"omgren sphere radius: about
150 pc for an O8 star in medium with density 0.2 cm$^{-3}$, twice that
for an O5 star. It is my suspicion that the concentration of OB
field stars in the general areas near HII regions would give rise to
similar spectral characteristics for the WIM as the leaking HII region
model.

\begin{figure}[htb]
\centering \leavevmode
\epsfxsize=0.50\textwidth
\epsfbox{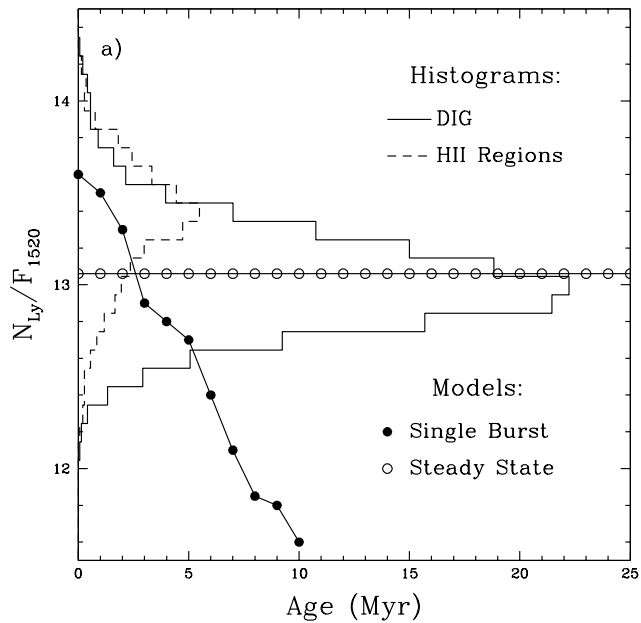}
\caption{Ratio of the Lyman continuum to far-UV (1520\AA)
  luminosity in HII regions and DIG (=WIM) regions in M33, compared to
  models from Hill et al. (1995). The Lyman continuum luminosity is
  inferred from the H$\alpha$ luminosity.  Note that the average ratio
  for the WIM is consistent with that predicted for a steady state
  star formation rate in a disk (from Hoopes \& Walterbos 1997). }
\label{fig-3}
\end{figure}

We are addressing this problem by analyzing far-UV images of the
stellar light, obtained with the Ultraviolet Imaging Telescope on the
ASTRO-1 and ASTRO-2 missions, in conjunction with the H$\alpha$ images
of the WIM. We test if the far-UV to H$\alpha$ intensity ratios across
galactic disks are consistent with those expected from luminous stars.
An example is shown in Figure~\ref{fig-3}.
The data appear consistent with ionization of the WIM by field stars.
However, there are several complications. Extinction effects are
troublesome in analyzing far-UV data. Some of the far-UV light in
regions of WIM could be due to light scattered from OB stars inside
HII regions. Also, while models such as shown in Figure~\ref{fig-3}
predict sufficient Lyman continuum output to ionize the WIM, we need
to determine if the {\it ionizing} stars are actually present. We are
doing this by analyzing HST far-UV images of selected regions in
nearby galaxies. Finally, the ionization stage of Helium is also
critical in testing the viability of field stars as the source of
ionization.

%
% Add as many section titles/contents as required.
%
% If you have subsections then use the
% \subsection{SUBSECTION TITLE}
% command and if you have subsubsections then use the
% \subsubsection{SUBSUBSECTION TITLE}
% command.  To use these commands, 
% first remove the % from the start of the line.

% It is preferable to embed your figures in the text. 
% One way to do this is to use the psfig style file and use the following
% commands to include the figures:

% \begin{figure}
% \begin{center}
% \psfig{file=filename.ps,height=10cm}
% \caption{Write your figure caption here.}
% \label{figlabel}            % for cross-references
% \end{center}
% \end{figure}

% To use the above commands, first remove the % from the beginning of
% the lines and then fill in your own values etc as appropriate.

% Tables
% Please consult previous issues of PASA
%  to see how tables are to be formatted.

\section*{Acknowledgements}

% Place acknowledgements here. Omit above \section command if there
% are no acknowledgements.
I appreciate the financial support from the LOC. The dedicated help of
Bruce Greenawalt, Charles Hoopes, Dave Thilker, and Vanessa Galarza at
NMSU, and Robert Braun from the NFRA is gratefully acknowledged.
Research supported by grants from NASA (NAG5-2426), the NSF
(AST-9123777 and AST-9617014) and a Cottrell Scholarship Award from
Research Corporation.

\section*{References}

% PASA uses the same conventions as ApJ for journal abbreviations.  Sample
% references are as follows. 
% Please follow the same format for your references.

%\reference Author, A.B. 1990 PASA 7, 2, 350

% for a journal article, or

% \reference Author, A.B. 1990 in This Is A Book Title, ed. Editor, C.D.,
% This Is A Publishers Name, 437

% for a book.
\reference Domg\"orgen, H., \& Mathis, J.S. 1994 Apj 428, 647
\reference Ferguson, A.M.N., Wyse, R.F.G., Gallagher, J.S., 
 Hunter, D.A. 1996a AJ 111, 226
\reference Ferguson, A.M.N., Wyse, R.F.G., Gallagher, J.S., 
 1996b AJ 112, 256
\reference Greenawalt, B.E., 1997 Ph.D. Thesis, New Mexico State University
\reference Greenawalt, B., Walterbos, R.A.M., \& Braun, R. 1997 ApJ 483, 666
\reference Hill, J.K. et al. 1995 ApJ 438, 181
\reference Hoopes, C.G., Walterbos, R.A.M., \& Greenawalt, B.E. 1996 
AJ 112, 1429
\reference Hoopes, C.G., \& Walterbos, R.A.M. 1997 in The Ultraviolet 
Universe at Low and High Redshift, Ed. W.H. Waller, AIP, in press
\reference Hunter, D.A., \& Gallagher, J.S. 1990 ApJ 362, 480
\reference Hunter, D.A., \& Gallagher, J.S. 1992 ApJ 391, L1
\reference Hunter, D.A., \& Gallagher, J.S. 1997 ApJ 475, 65
\reference Kulkarni, S.R., \& Heiles, C. 1988 in Galactic and Extragalactic
Radio Astronomy, Eds G. Verschuur \& K.I. Kellermann, 95
\reference Rand, R.J. 1997, ApJ 474, 129
\reference Reynolds, R.J. 1988 ApJ 333, 341
\reference Reynolds, R.J., \& Tufte, S.L. 1993 ApJ 439, L17
\reference Slavin, J.D., Shull, J.M, \& Begelman, M.C. 1993 ApJ 407, 83
\reference Walterbos, R.A.M., \& Braun, R. 1992 A\&AS 92, 625
\reference Walterbos, R.A.M., \& Braun, R. 1994 ApJ 431, 156
 
% Add as many references as required.

\end{document}